# Development of Imaging Calorimetry[1]

Burak Bilki, José Repond, Lei Xia
Argonne National Laboratory
9700 S. Cass Avenue, Argonne, IL 60439, U.S.A.

**Abstract:** This paper will argue for continued effort in developing imaging calorimeters for future colliders and/or upgrades to existing detectors. Imaging calorimeters offer a plethora of advantages beyond their application in conjunction with Particle Flow Algorithms. Further R&D is needed to turn the first generation prototypes into viable detectors for colliding beam experiments.

1. **What are imaging calorimeters?**

Imaging calorimeters are calorimeters with a finely granulated readout. Instead of the traditional calorimeter towers or bulky crystals, each connected to a single readout channel, the readout of imaging calorimeters is segmented into small areas, both longitudinally and laterally. As an example, the DHCAL [1] uses Resistive Plate Chambers (RPCs) as active elements, which are read out with an array of 1 x 1 $cm^2$ pads. The fine segmentation of the DHCAL is illustrated in Fig. 1 with the event display of a 16 GeV pion. In general, imaging calorimeters are based on the sandwich design, with alternating active and passive elements (absorbers). At least in principle, finely segmented crystal (total absorption) calorimeters are also conceivable as imaging calorimeters.

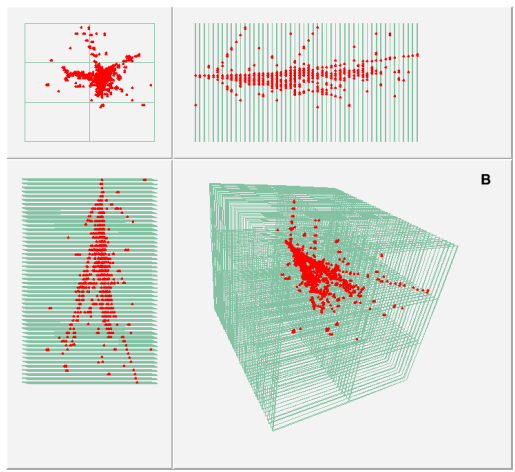

**Figure 1.** Event display of a 16 GeV pion in the DHCAL. The absorber was minimal and consisted only of 2 mm Copper and 2 mm Steel per layer.

---

[1] White paper for the Snowmass 2013 proceedings

The fine segmentation of the readout, naturally, leads to a large number of readout channels. The segmentation also implies that the front-end electronic be placed on the surface of the active element and is thus embedded into the calorimeter structure. The large number of channels renders the calibration of the device, in particular the equalization of the response of each channel, a formidable challenge.

In the past decade, several technologies have been explored as candidates for the active element of imaging calorimeters, all based on the sandwich design: silicon wafers, scintillator strips or pads, RPCs, Gas Electron Multipliers (GEMs), and Micromegas.

## 2. Advantages of imaging calorimeters

Imaging calorimeters offer a number of distinct advantages. Foremost is their usefulness for the application of Particle Flow Algorithms (PFAs). PFAs attempt to measure each particle in a hadronic jet individually with the component providing the best momentum/energy resolution. Thus charged particles are measured with a precision tracker, photons with an electromagnetic calorimeter and the remaining neutral hadrons (neutrons and $K_L^0$) are measured with the electromagnetic and hadronic calorimeter combined. To date PFAs have been successfully applied to detectors which have not been optimized for their application, such as CDF, ZEUS, and CMS [2]. However, detailed Monte Carlo simulation have shown [3], that detectors optimized for PFAs can perform significantly better than existing detectors. Such detectors feature finely segmented imaging calorimeters.

With finely segmented calorimeters, the measurement of individual hadronic jets is greatly improved. Furthermore, in multijet events the assignment of calorimeter energy deposits to the individual jets of the event is facilitated, leading to a better resolution on the mass of dijets. Detailed Monte Carlo simulations have shown that hadronic decays of the W and Z bosons can be reconstructed with 3 – 4% accuracy in a large energy range [3, 4].

Most sandwich calorimeters are not compensating and feature an electron over hadron ratio e/h > 1. Since the fine segmentation of imaging calorimeters can be used to identify the electromagnetic subshowers in a hadron shower, the response can be made to be compensating by applying so-called software compensation techniques. These weigh the subshowers appropriately to achieve a de facto e/h close to unity. In the process, the energy resolution for single hadrons is improved by about 25% [5].

Energy leaking from the back of calorimeters typically results in low-energy tails in the energy response, thus degrading the energy resolution. Sophisticated algorithms using the information available in imaging calorimeters can correct for leakage and, therefore, improve the energy resolution for single particles [6]. Furthermore, if the calorimeter is placed in a magnetic field, the curvature of tracks exiting the calorimeter can be measured and, thus, their energy recovered (this method has not yet been explored in detail.)

With an imaging electromagnetic calorimeter, the two photons from neutral pion decays can be measured individually, up to high energies. Using the neutral pion mass as a constraint can further improve the hadronic jet energy. But even more importantly, with reconstructed neutral pions, the polarization of $\tau^- \rightarrow \rho^- \nu_\tau \rightarrow \pi^- \pi^0 \nu_\tau$ in a LC environment can be measured.

Imaging calorimeters give a unprecedented insight into the spatial development of hadronic showers. Measurements of the location of the first hadronic interaction, of the longitudinal and transverse shapes can be performed with great precision [7, 8]. Measurements, such as these, are essential for the validation of the various hadronic shower models currently being offered within the GEANT4 framework.

### 3. Status of the R&D with imaging calorimeters

The last decade has seen great progress in studying and understanding imaging calorimeters. A number of large prototypes were designed, built and tested in test beams. Even though the final results are still outstanding, technologies, such as Silicon pads, scintillator pads/strips, RPCs and Micromegas have all been evaluated and their viability as active elements have been proven.

Most of the work on imaging calorimetry has been conducted under the auspices of the international CALICE collaboration [9]. The notable exception is the development of a Silicon-pad ECAL within the framework of the SiD concept [10]. The work by CALICE and SiD is directed at future Linear Colliders (LCs): the International Linear Collider (ILC) [11] and the Compact Linear Collider (CLIC) [12]. In the following the major projects in this endeavor are briefly presented:

The concept of the CALICE Silicon Tungsten ECAL was extensively tested with a small-scale prototype [13]. This calorimeter contained 30 layers, read out with 1 x 1 $cm^2$ pads. The signals from the 9,720 individual readout channels were digitized with electronics placed off the detector. The detector was exposed to particle beams at DESY, CERN and Fermilab.

The SiD ECAL group is currently assembling a 30 layer structure with Silicon-pads and Tungsten absorber plates. The pads are shaped as hexagons with an area of 0.13 $cm^2$ each. The readout is based on the KPiX chip, which is located directly on the wafer. Tests in the SLAC test beam are planned for the next few months.

The CALICE Scintillator-Tungsten ECAL features scintillator strips, read out with MPPCs [14]. In subsequent layers the strips, each with an area of 1.0 x 4.5 $cm^2$, were oriented alternatively in the x or y direction. A small-scale prototype with 30 layers was exposed to the DESY and Fermilab test beams.

The CALICE Scintillator Analog Hadron Calorimeter (AHCAL) uses Scintillator pads coupled to Silicon Photomultipliers (SiPMs) and a multi-bit readout as active media [15]. A large prototype calorimeter was built, containing 38 layers with a total channel count of close to 8,000. The scintillator pads range from 3 x 3 $cm^2$ in the core of the stack to 6 x 6 and 12 x 12 $cm^2$ on the outskirts of a given layer or in the layers located towards the back of the calorimeter. The AHCAL was the first large detector to make extensive use of SiPMs. It was extensively tested at DESY, CERN and FNAL with both Steel (Fe-AHCAL) and Tungsten (W-AHCAL) absorber plates. Figure 2 shows a photograph of a layer before being optically sealed.

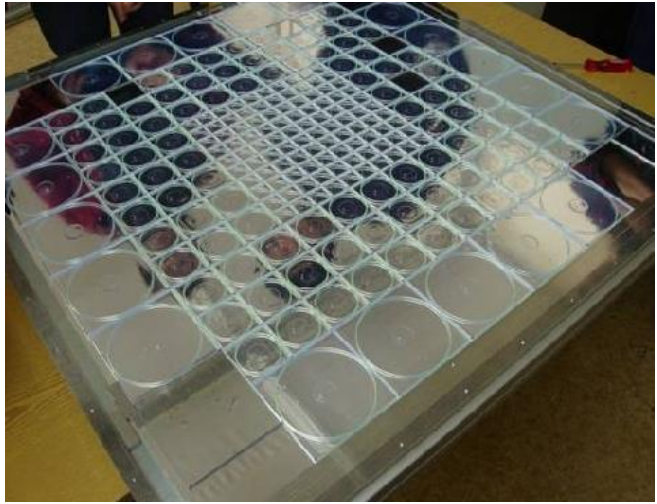

**Figure 2.** Photograph of a layer of the AHCAL prototype, before being optically sealed.

The CALICE Digital Hadron Calorimeter (DHCAL) consisted of up to 54 1 x 1 $m^2$ layers of RPCs. The readout was segmented into 1 x 1 $cm^2$ pads, each read out with a 1-bit (or digital) resolution. For the first time in calorimetry, the front-end electronics (including the digitization) was embedded into the structure itself. With 54 layers, this prototype counted close to 500,000 readout channels, a world record in both calorimetry and for RPC systems. It was tested at FNAL and CERN with both Steel (Fe-DHCAL) and Tungsten (W-DHCAL) absorber plates.

The CALICE Semi-Digital Hadron Calorimeter (SDHCAL) also used 1 x 1 $m^2$ layers of RPCs as active media. The readout was segmented into 1 x 1 $cm^2$ pads, but here read out with a 2-bit resolution, corresponding to the application of three distinct thresholds. The front-end electronics was embedded into the calorimeter structure and was successfully operated in power pulsed mode to reduce the generation of heat and the need for active cooling. The prototype was tested at CERN with Steel absorber plates.

The CALICE Micromegas group tested several 1 x 1 $m^2$ layers in test beams at CERN and DESY. The readout system was similar to the one employed by the SDHCAL.

4. Opportunities and challenges for imaging calorimeters at the LHC and LCs

The current forward calorimeters of the two large LHC experiments, CMS and ATLAS, are not expected to be adequate for the conditions imposed by the planned high-luminosity LHC running. Also, with the successful application of PFAs to the analysis of CMS data [2], it appears worthwhile to consider replacing the existing forward calorimeters with imaging calorimeters. The forward regions at the LHC, however, entail a number of challenges. Foremost are the expected particle rates, up to 50 MHz in the most forward parts of the calorimeter. The active elements of such calorimeters are required to provide a stable response independent of particle rates, necessitating detailed tests in high-rate environments, e.g. at the GIF++ facility at CERN [16]. It is worthwhile to note, that these particle rates eliminate e.g. current RPCs with either glass or Bakelite resistive plates as active elements. To remedy this, low-resistivity glass is being developed by COE college (Iowa) [17]. This novel glass is expected to increase the rate capability of RPCs by several orders of magnitude.

The accumulated dose expected in 10 years of high-luminosity running at the LHC is staggering with estimated fluences of $10^{16}$ neutrons/cm$^2$ in certain regions of the detector. This necessitates the development of radiation hard active elements, as well as radiation hard front-end electronics.

With the LHC colliding at a rate of 40 MHz and pile-ups of up to 140 collisions per crossing the signals of the upgraded forward calorimeter need to feature fast decay times to avoid shifts in the baseline as function of collision rate. Also, a timing resolution of better than 25 ns is desirable.

Concerning the LCs, most concerns, such as rate capability and segmentation have been addressed by the existing prototypes listed in section 3. At the ILC, due to the low bunch-crossing rate, exact timing measurements are not necessary. On the other hand, in order to suppress 2-$\gamma$ backgrounds at CLIC energies, a timing resolution of the order of 1 ns is mandatory [4].

5. **Further R&D with imaging calorimeters**

The major thrust of the ongoing R&D for imaging calorimeters is addressing the remaining technical issues related to the design of LC detectors and developing a viable option for the LHC forward calorimeters.

Several active elements are being considered for the LHC upgrade, such as low-resistivity RPCs, GEMs, Micromegas, and Secondary Emission Calorimeters (SECs). These are expected to provide the necessary performance for the HCAL, being able to handle the high rates and the integrated radiation. The choices of active elements for the ECAL, on the other hand, are not obvious: can the same technologies be extended into the ECAL? Given the high rates and particle energies, will these devices still provide a linear response to electromagnetic showers? A comprehensive test program is being planned to address these issues.

After the highly successful program related to the large scale prototypes, the CALICE collaboration moved on to address the remaining technical issues for calorimetry at a LC with the design, building and testing of so-called technological prototypes. The Silicon-Tungsten and Scintillator-Tungsten ECAL groups are testing several layers with integrated electronics, which are inserted into 'realistic' absorber structures. The AHCAL also assembled several layers with integrated electronics which are now being tested in the DESY test beam.

Imaging calorimeters require sophisticated distribution systems for high-voltage (HV), low-voltage (LV) and (for gaseous detectors) gas. These systems are envisaged to distribute the HV/LV/gas from a primary supply to a large number of channels, e.g. the individual chambers in a module. The distribution system is expected to be able to adjust e.g. the HV (within a reasonable range), to turn on and off individual channels and to monitor the voltage and current of each channel. Development of such distribution systems has begun, but is in need of further funding.

Finally, specifically for the operation of RPCs, and possibly also for other Micro-pattern gas detectors, a gas recirculation system is mandatory, to both protect the environment from greenhouse gases and to reduce the operational cost. Recirculation systems for RPCs are currently operational for the LHC detectors at CERN. Further R&D is needed for gaseous detectors optimized for use in calorimeter systems.

## 6. Conclusions

Imaging calorimeters offer a plethora of advantages compared to traditional calorimeters. Their development has been spear-headed by the CALICE collaboration and the SiD ECAL group. Several small- and large-scale prototypes have been built to demonstrate the viability of the technique.

Imaging calorimeters are being considered for the upgrade of the forward calorimeters at the LHC and for detectors at future Lepton Colliders. In both cases, additional effort is needed to address the remaining technical issues. In case of the LHC forward calorimeters, a viable active element, capable of sustaining the high particle rates and radiation doses, needs to be identified.

The development of imaging calorimetry requires a long-term effort. It will eventually lead to a significant extension of the physics reach of colliding beam detectors.